\documentclass[trackchanges,twocolumn]{aastex701}

\usepackage{amsmath}
\usepackage{comment}
\usepackage{multirow} 
\usepackage{array}    
\usepackage{booktabs} 

\begin{document}

\title{Enhanced Mid-Photospheric Heating Around a Solar Pore}

\author[orcid=0000-0002-0570-4029]{Ana Bel\'en Gri\~n\'on-Mar\'in}
\affiliation{Institute for Solar Physics, Department of Astronomy, Stockholm University, \\ Albanova University Centre, SE-106 91 Stockholm, Sweden.}
\email[show]{ana.grinonmarin@astro.su.se}  

\author[orcid=0000-0003-2359-9039]{Adur Pastor Yabar} 
\affiliation{Institute for Solar Physics, Department of Astronomy, Stockholm University, \\ Albanova University Centre, SE-106 91 Stockholm, Sweden.}
\email{adur.pastor@astro.su.se}

\author[orcid=0000-0002-4640-5658]{Jaime de la Cruz Rodr\'iguez} 
\affiliation{Institute for Solar Physics, Department of Astronomy, Stockholm University, \\ Albanova University Centre, SE-106 91 Stockholm, Sweden.}
\email{jaime@astro.su.se}

\author[orcid=0000-0002-3418-8449,sname='Solanki']{Sami~K.~Solanki} \affiliation{Max-Planck-Institut für Sonnensystemforschung, Justus-von-Liebig-Weg 3, 37077 Göttingen, Germany}\email{solanki@mps.mpg.de}		
		
\author[orcid=0000-0003-1459-7074,sname='Lagg']{Andreas~Lagg} \affiliation{Max-Planck-Institut für Sonnensystemforschung, Justus-von-Liebig-Weg 3, 37077 Göttingen, Germany}\email{lagg@mps.mpg.de}		
\author[orcid=0000-0002-9972-9840,sname='Gandorfer']{Achim~Gandorfer} \affiliation{Max-Planck-Institut für Sonnensystemforschung, Justus-von-Liebig-Weg 3, 37077 Göttingen, Germany}\email{gandorfer@mps.mpg.de}		
\author[orcid=0000-0002-3387-026X,sname='del~Toro~Iniesta']{Jose~Carlos~del~Toro~Iniesta} \affiliation{Instituto de Astrofísica de Andalucía, CSIC, Glorieta de la Astronomía s/n, 18008 Granada, Spain}\affiliation{Spanish Space Solar Physics Consortium}\email{jti@iaa.es}		
\author[orcid=0000-0002-5054-8782,sname='Katsukawa']{Yukio~Katsukawa} \affiliation{National Astronomical Observatory of Japan, 2-21-1 Osawa, Mitaka, Tokyo 181-8588, Japan}\affiliation{Department of Earth and Planetary Science, The University of Tokyo, 7-3-1, Hongo, Bunkyo-ku, Tokyo 113-0033, Japan}\affiliation{Department of Astronomical Science, The Graduate University for Advanced Studies (SOKENDAI), 2-21-1 Osawa, Mitaka, Tokyo 1818588, Japan}\email{yukio.katsukawa@nao.ac.jp}		
\author[orcid=0000-0002-0787-8954,sname='Bernasconi']{Pietro~Bernasconi} \affiliation{Johns Hopkins University Applied Physics Laboratory, 11100 Johns Hopkins Road, Laurel, Maryland, USA}\email{pietro.bernasconi@jhuapl.edu}		
\author[sname='Berkefeld']{Thomas~Berkefeld} \affiliation{Institut für Sonnenphysik (KIS), Georges-Köhler-Allee 401a, 79110 Freiburg, Germany}\email{thomas.berkefeld@leibniz-kis.de}		
\author[orcid=0009-0009-4425-599X,sname='Feller']{Alex~Feller} \affiliation{Max-Planck-Institut für Sonnensystemforschung, Justus-von-Liebig-Weg 3, 37077 Göttingen, Germany}\email{feller@mps.mpg.de}		
\author[orcid=0000-0001-6317-4380,sname='Riethmüller']{Tino~L.~Riethmüller} \affiliation{Max-Planck-Institut für Sonnensystemforschung, Justus-von-Liebig-Weg 3, 37077 Göttingen, Germany}\email{riethmueller@mps.mpg.de}

\author[orcid=0000-0001-9228-3412,sname='Álvarez-Herrero']{Alberto~Álvarez-Herrero} \affiliation{Instituto Nacional de T\'ecnica Aeroespacial (INTA), Ctra. de Ajalvir, km. 4, E-28850 Torrejón de Ardoz, Spain}\affiliation{Spanish Space Solar Physics Consortium}\email{alvareza@inta.es}		
\author[orcid=0000-0001-5616-2808,sname='Kubo']{Masahito~Kubo} \affiliation{National Astronomical Observatory of Japan, 2-21-1 Osawa, Mitaka, Tokyo 181-8588, Japan}\email{masahito.kubo@nao.ac.jp}		
\author[orcid=0000-0003-3490-6532,sname='Smitha']{H.~N.~Smitha} \affiliation{Max-Planck-Institut für Sonnensystemforschung, Justus-von-Liebig-Weg 3, 37077 Göttingen, Germany}\email{narayanamurthy@mps.mpg.de}		
\author[orcid=0000-0001-8829-1938,sname='Orozco~Suárez']{David~Orozco~Suárez} \affiliation{Instituto de Astrofísica de Andalucía, CSIC, Glorieta de la Astronomía s/n, 18008 Granada, Spain}\affiliation{Spanish Space Solar Physics Consortium}\email{orozco@iaa.es}		
\author[sname='Grauf']{Bianca~Grauf} \affiliation{Max-Planck-Institut für Sonnensystemforschung, Justus-von-Liebig-Weg 3, 37077 Göttingen, Germany}\email{grauf@mps.mpg.de}		
\author[sname='Carpenter']{Michael~Carpenter} \affiliation{Johns Hopkins University Applied Physics Laboratory, 11100 Johns Hopkins Road, Laurel, Maryland, USA}\email{michael.carpenter@jhuapl.edu}		
\author[sname='Bell']{Alexander~Bell} \affiliation{Institut für Sonnenphysik (KIS), Georges-Köhler-Allee 401a, 79110 Freiburg, Germany}\email{albe@leibniz-kis.de}		
\author[orcid=0000-0001-7764-6895,sname='Martínez~Pillet']{Valentín~Martínez~Pillet} \affiliation{Instituto de Astrofísica de Canarias, Vía Láctea, s/n, E-38205 La Laguna, Spain}\affiliation{Spanish Space Solar Physics Consortium}\email{vmpillet@iac.es}

\author[orcid=0000-0002-7318-3536,sname='Bailén']{Francisco~Javier~Bailén} \affiliation{Instituto de Astrofísica de Andalucía, CSIC, Glorieta de la Astronomía s/n, 18008 Granada, Spain}\affiliation{Spanish Space Solar Physics Consortium}\email{fbailen@iaa.es}		
\author[orcid=0000-0002-2055-441X,sname='Blanco~Rodríguez']{Julian~Blanco~Rodríguez} \affiliation{Universitat de Valencia Catedrático José Beltrán 2, E-46980 Paterna-Valencia, Spain}\affiliation{Spanish Space Solar Physics Consortium}\email{julian.blanco@uv.es}		
\author[orcid=0000-0003-4319-2009,sname='Castellanos~Durán']{Juan~Sebastián~Castellanos~Durán} \affiliation{Max-Planck-Institut für Sonnensystemforschung, Justus-von-Liebig-Weg 3, 37077 Göttingen, Germany}\email{castellanos@mps.mpg.de}		
\author[orcid=0009-0002-6808-5154,sname='Harnes']{Edvarda~Harnes} \affiliation{Max-Planck-Institut für Sonnensystemforschung, Justus-von-Liebig-Weg 3, 37077 Göttingen, Germany}\email{harnes@mps.mpg.de}		
\author[orcid=0000-0001-6029-7529,sname='Hoelken']{Johannes~Hoelken} \affiliation{Max-Planck-Institut für Sonnensystemforschung, Justus-von-Liebig-Weg 3, 37077 Göttingen, Germany}\email{hoelken@mps.mpg.de}		
\author[orcid=0000-0003-1409-1145,sname='Iglesias']{Francisco~A.~Iglesias} \affiliation{Max-Planck-Institut für Sonnensystemforschung, Justus-von-Liebig-Weg 3, 37077 Göttingen, Germany}\affiliation{Grupo de Estudios en Heliofísica de Mendoza, CONICET, Universidad de Mendoza, Boulogne sur Mer 683, 5500 Mendoza, Argentina}\email{iglesias@mps.mpg.de}		
\author[orcid=0000-0002-4669-5376,sname='Ishikawa']{Ryohtaroh~T.~Ishikawa} \affiliation{National Institute for Fusion Science, 322-6 Oroshi-cho, Toki City 509-5292, Japan}\email{ishikawa.ryohtaro@nifs.ac.jp}		
\author[orcid=0000-0001-7452-0656,sname='Kawabata']{Yusuke~Kawabata} \affiliation{National Astronomical Observatory of Japan, 2-21-1 Osawa, Mitaka, Tokyo 181-8588, Japan}\email{kawabata.yusuke@nao.ac.jp}		
\author[orcid=0000-0002-1043-9944,sname='Matsumoto']{Takuma~Matsumoto} \affiliation{Centre for Integrated Data Science, Institute for Space-Earth Environmental Research, Nagoya University, Furocho, Chikusa-ku, Nagoya, Aichi 464-8601, Japan}\email{takuma.matsumoto@gmail.com}		
\author[orcid=0000-0002-7044-6281,sname='Oba']{Takayoshi~Oba} \affiliation{Max-Planck-Institut für Sonnensystemforschung, Justus-von-Liebig-Weg 3, 37077 Göttingen, Germany}\affiliation{National Astronomical Observatory of Japan, 2-21-1 Osawa, Mitaka, Tokyo 181-8588, Japan}\email{oba@mps.mpg.de}		
\author[orcid=0000-0003-0175-6232,sname='Siu-Tapia']{Azaymi~L.~Siu-Tapia} \affiliation{Instituto de Astrofísica de Andalucía, CSIC, Glorieta de la Astronomía s/n, 18008 Granada, Spain}\affiliation{Spanish Space Solar Physics Consortium}\email{siu@iaa.es}		
\author[orcid=0000-0003-1483-4535,sname='Strecker']{Hanna~Strecker} \affiliation{Instituto de Astrofísica de Andalucía, CSIC, Glorieta de la Astronomía s/n, 18008 Granada, Spain}\affiliation{Spanish Space Solar Physics Consortium}\email{streckerh@iaa.es}		
\author[orcid=0000-0003-1971-5551,sname='Vukadinović']{Dušan~Vukadinović} \affiliation{Institut für Physik, Universität Graz, Universitätsplatz 5, 8010 Graz, Austria}\affiliation{Max-Planck-Institut für Sonnensystemforschung, Justus-von-Liebig-Weg 3, 37077 Göttingen, Germany}\email{vukadinovic@mps.mpg.de}

\author[orcid=0000-0001-5686-3081,sname='Hara']{Hirohisa~Hara} \affiliation{National Astronomical Observatory of Japan, 2-21-1 Osawa, Mitaka, Tokyo 181-8588, Japan}\email{hirohisa.hara@nao.ac.jp}		
\author[orcid=0000-0003-4764-6856,sname='Shimizu']{Toshifumi~Shimizu} \affiliation{Department of Earth and Planetary Science, The University of Tokyo, 7-3-1, Hongo, Bunkyo-ku, Tokyo 113-0033, Japan}\affiliation{Institute of Space and Astronautical Science, Japan Aerospace Exploration Agency, 3-1-1, Yoshinodai, Chuo-ku, Sagamihara, Kanagawa 252-5210, Japan}\email{shimizu.toshifumi@isas.jaxa.jp}		


\begin{abstract}

\begin{center} 
  \small\itshape Accepted for publication in The Astrophysical Journal Letters. 
\end{center}

We present depth-stratified atmospheric reconstructions of a solar pore observed with the Sunrise Chromospheric Infrared SpectroPolarimeter (SCIP) aboard \textsc{Sunrise~iii}. Using spatially coupled NLTE spectropolarimetric inversions, we characterize the thermal, magnetic, and velocity structure of the pore and its surroundings from the photosphere to the chromosphere over approximately two hours. The inversions recover the canonical photospheric properties of pores: a cool umbra with strong magnetic fields dominated by the vertical component. Most photospheric velocities are subsonic, but localized downflows near the pore boundary reach transonic or potentially supersonic values around $\log\tau_{500}=0$. At higher atmospheric layers, the thermal contrast between the pore and its surroundings decreases substantially. We also detect localized mid-photospheric temperature enhancements forming a patchy, ring-like structure in the wings of the Ca\,{\sc ii} 8542\,{\AA} line, without a clear continuum counterpart. The temperature excess reaches approximately $300$\,K and corresponds to an excess thermal-energy density of up to $170$\,J\,m$^{-3}$. These regions coincide with an enhanced transverse magnetic-field component and the expanding magnetic canopy surrounding the pore. The structure is detected in all three rasters, although the cadence does not establish whether individual patches persist or repeatedly re-form. While the heating mechanism cannot be uniquely identified, the inferred magnetic topology supports an association with processes operating in the expanding canopy. These results demonstrate the value of pores for investigating the coupling between magnetic topology, convection, and atmospheric energy transport.

\end{abstract}




\section{Introduction} 
\label{intro}

In the photosphere, pores appear as small, dark magnetic features compared to sunspots, composed of only an umbral core without a penumbral structure; they have an average area of about $4\, {\rm Mm}^{2}$ \citep{verma2014} and lifetimes ranging from a few hours to just over one day \citep{tlatov2023}. The magnetic field in pores is strong and mostly vertical, typically reaching values of the order of kilo-Gauss. Pores harbor magnetic fluxes of the order of $10^{20}$ Mx lying in between sunspots ($10^{20}-10^{22}$ Mx) and magnetic elements composing \emph{plage} ($10^{18}$ Mx) \citep{simon1970,hurlburt2000,thomas2008}. Due to their strong magnetic character and their different structure compared to fully developed sunspots, pores are often considered ideal laboratories for studying the interaction between magnetic fields and convection in the solar atmosphere \citep{sobotka1999}. Unlike fully developed sunspots, pores lack a photospheric penumbra and are generally described as magnetic structures dominated by strong, predominantly vertical fields \citep{simon1970,rucklidge1995,rempel2009}.

The appearance of pores in the photosphere and the chromosphere is clearly different. The magnetic field of pores expands with height \citep{sobotka2013,lindner2023}, leading to the formation of a canopy-like configuration in the upper photosphere and chromosphere, where the field becomes more inclined and spreads over the surrounding photosphere. In some cases, this configuration is accompanied by a filamentary superpenumbral structure, as reported by \citet{sobotka2013} from \ion{Ca}{2} IR observations, similar to superpenumbrae around sunspots. This behavior is a result of the decrease of gas pressure with height, which can enhance the relative importance of the magnetic field in shaping plasma dynamics and facilitate the formation of filamentary structures. Such magnetic configurations may create favorable conditions for localized energy dissipation in the upper atmosphere above the pore and its immediate surroundings, although the dominant mechanisms remain uncertain \citep[e.g.,][]{khomenko2012,martinez2017}.

Previous studies, based on photospheric observations of solar pores, commonly reveal weak or suppressed flows in their interiors, and narrow annular or localized downflows at their boundaries \citep{keil1999,tritschler2002,sankarasubramanian2003,hirzberger2003,giordano2008,sobotka2012,kitiashvili2013,cho2013,bharti2016,ermolli2017}. These studies have reported photospheric velocities ranging from 300 m\,s$^{-1}$ to 2.5 km\,s$^{-1}$ at the outer boundary of pores. Radiative-MHD simulations also reproduce strong downflow lanes around pore-like magnetic concentrations and emphasize their role in concentrating and redistributing magnetic flux \citep{stein2000,cameron2007,kitiashvili2013}. In the chromosphere, much larger and occasionally supersonic downflows have been associated with magnetic-loop drainage, siphon flows, and shocks \citep{uitenbroek2006,lagg2007,dubey2026}.

In this study, we investigate the thermal, magnetic, and velocity structure of a solar pore and its surroundings using depth-stratified Non-Local Thermodynamic Equilibrium (NLTE) inversions of simultaneous photospheric and chromospheric spectropolarimetric observations. Particular attention is given to a localized mid-photospheric temperature enhancements around the pore boundary.

\renewcommand{\thefootnote}{\fnsymbol{footnote}}
\setcounter{footnote}{0}


\section{Observations and Methodology}
\label{obs&met}

\subsection{Data}
\begin{figure*}
 \centering
 \includegraphics{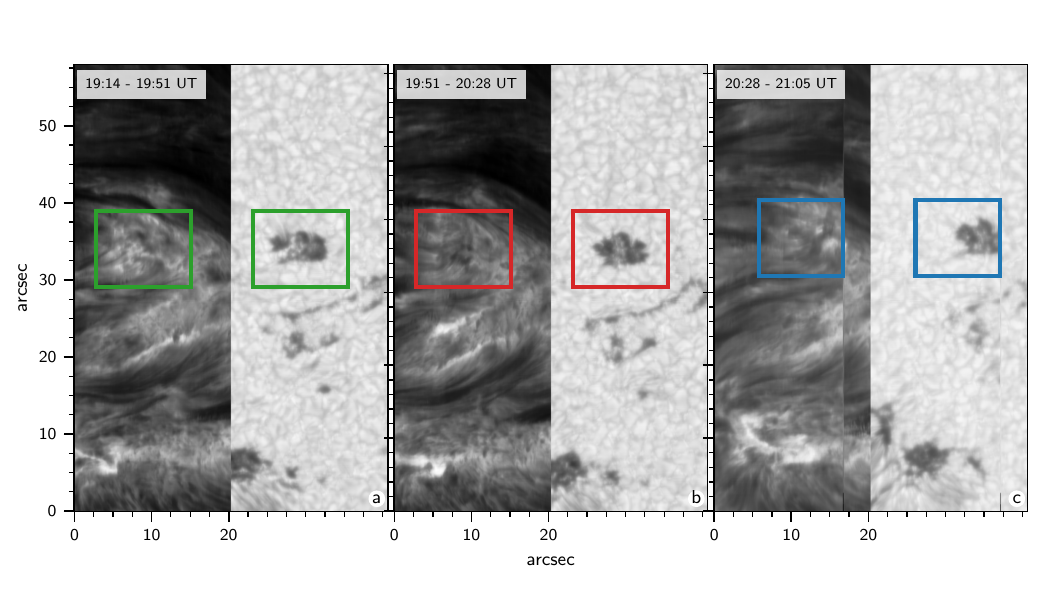}
 \caption{SCIP observations of the solar pore studied in this work. The panels show the \ion{Ca}{2} 8542\,\AA\ line-core intensity (on the left) and continuum intensity (on the right) for the three consecutive rasters (see top-left labels for the observation time). The color squares mark the field of view on which our study has focused.}
 \label{fig:observed_fov}
\end{figure*}

The balloon-borne solar observatory \textsc{Sunrise iii} \citep{solanki2010_sunrise,barthol2011_sunrise,solanki2017_sunrise}, during its third successful flight \textsc{Sunrise~iii} \citep{lagg2025_sunrise,solanki2026_sunrise}, observed a flux emergence region on 2024 July 10, belonging to the active region NOAA AR13738, which was located around X=-60$^{\prime\prime}$ and Y=-200$^{\prime\prime}$, corresponding to $\mu=0.97$. The observation has \textsc{Sunrise iii} ID  02\_EMEF (see \citealt{solanki2026_sunrise}). In particular, we have used the three rasters recorded by the Sunrise Chromospheric Infrared SpectroPolarimeter (SCIP; \citealt{katsukawa2026_scip}) using Channel 1 (8466 {\AA} - 8547 {\AA}) in full spectropolarimetric mode (the data were reduced and provided by the SCIP team). The three rasters started at 19:14, 19:51, and 20:28 UT, corresponding to a start-to-start cadence of approximately 37 min. Each raster occupied nearly the full interval between consecutive start times, with the final raster ending at 21:05 UT. The observed spectral range includes spectral lines that are sensitive to magnetic fields in the solar photosphere and chromosphere (see the upper part of Table \ref{tab:spectralLines}) with a spectral sampling of 39.5 m\AA. The integration time was 10 seconds per slit position and the scanning step size and the image scale along the slit were $0^{\prime\prime}.094$. Each raster covered a field of view (FOV) of approximately $60^{\prime\prime} \times 20^{\prime\prime}$ (see Fig. \ref{fig:observed_fov}\footnote{A pointing jump during the third raster resulted in incomplete spatial coverage of the pore. Only the valid observed portion of the raster was included in the analysis.}). In the following, we focused on the areas highlighted with colored squares in this figure.

\begin{table}[]
  \caption{Parameters used in the inversions. The upper part of the table lists the atomic parameters of the inverted spectral lines, while the lower part gives the number of nodes and regularization parameters for each physical quantity in each inversion cycle (C1, C2, C3). Symbol $*$ on the Species highlights that collisional parameters with neutral hydrogen follow \cite{BPM}. The nodes are equally spaced in column mass between $\lg\xi>-7$ and $\lg\xi<1.45$ in units of $\lg g\,cm^{-2}$.}
  \label{tab:spectralLines}    
  \centering
  \renewcommand{\arraystretch}{1.2} 
  {\footnotesize
  \begin{tabular}{c|ccc|ccc|ccc}
    \hline

    \multicolumn{2}{c}{\textbf{}} & \multicolumn{2}{c}{\textbf{Species}} & \multicolumn{2}{c}{\textbf{Wavel.}} & \multicolumn{2}{c}{} & \multicolumn{2}{c}{\textbf{Land\'e}} \\ 

    \multicolumn{2}{c}{\textbf{\#}} & \multicolumn{2}{c}{\textbf{ion.}} & \multicolumn{2}{c}{\textbf{({\AA})}} & \multicolumn{2}{c}{\textbf{$\lg{gf}$}} & \multicolumn{2}{c}{\textbf{Factor}} \\ \hline

    \multicolumn{2}{c}{1} & \multicolumn{2}{c}{\ion{Fe}{1}$*$}  & \multicolumn{2}{c}{8468.406} & \multicolumn{2}{c}{-1.817} & \multicolumn{2}{c}{2.49} \\
    
    \multicolumn{2}{c}{2} & \multicolumn{2}{c}{\ion{Fe}{1}}     & \multicolumn{2}{c}{8468.619} & \multicolumn{2}{c}{-1.234} & \multicolumn{2}{c}{1.74} \\
    
    \multicolumn{2}{c}{3} & \multicolumn{2}{c}{\ion{Ca}{2}$*$} & \multicolumn{2}{c}{8542.088} & \multicolumn{2}{c}{-0.476} & \multicolumn{2}{c}{1.10} \\ \hline

\hline
\hline

\multicolumn{1}{c|}{} & \multicolumn{3}{c|}{}      & \multicolumn{3}{c|}{\textbf{Lon. Reg.}} & \multicolumn{3}{c}{\textbf{Perp. Reg.}} \\
\multicolumn{1}{c|}{} & \multicolumn{3}{c|}{\textbf{Nodes}} & \multicolumn{3}{c|}{\textbf{Weights}}   & \multicolumn{3}{c}{\textbf{Weights}} \\

\textbf{Param.} & \textbf{C1} & \textbf{C2} & \textbf{C3} & \textbf{C1} & \textbf{C2} & \textbf{C3} & \textbf{C1} & \textbf{C2} & \textbf{C3} \\ \hline
$T$          & 5 & 9 & 15 & 1 & 10 & 1 & $10^{4}$ & $10^{3}$ & 160\\ 
$v_{los}$    & 3 & 6 & 11 & 1 & 10 & 5 & $10^{2}$ & 10 & 1\\ 
$v_{turb}$   & 2 & 5 & 7 & 10 & 10 & 8 & $10^{4}$ & $10^{3}$ & 40\\ 
$B_{los}$    & 2 & 3 & 7 & 0 & 10 & 4 & $10^{2}$ & 10 & 1\\ 
$B_{perp}$   & 1 & 2 & 5 & 0 & 10 & 2 & $10^{2}$ & 10 & 4\\ 
$\phi$       & 1 & 1 & 2 & 0 & 0 & 1 & 1 & 1 & 1\\ 

\end{tabular}
}
\end{table}

\subsection{Inversions}
\label{inversions}
We performed NLTE inversions using the multi-resolution spatially coupled STockholm inversion Code (STiC; \citealt{pastoryabar2026,delacruz2019_stic2}), to infer the atmospheric stratification of the temperature, the line-of-sight (LOS) velocity, the full magnetic field vector and the microturbulent velocity. This version of the code builds upon the original STiC \citep{delacruz2016_stic1,delacruz2019_stic2}, which is a suitable tool to characterize the NLTE line formation of the \ion{Ca}{2} 8542 {\AA} line. In the new version, this NLTE radiative transfer treatment is expanded with a better characterization of the multi-instruments/multi-resolution effects (including the effect of the point-spread function (PSF) in a similar fashion to the original development by \citealt{vannoort2012}) together with the inclusion of perpendicular to the LOS spatial regularization (within the same framework as in \citealt{delacruzrodriguez2019multi, delacruzrodriguez2024}). Because the target was observed near disk center, $B_{\rm los}$ and $B_{\rm perp}$ approximately represent the local vertical and horizontal components, respectively.

We included in the inversion two blended \ion{Fe}{1} spectral lines around \ion{Fe}{1} 8468.4 {\AA} \citep{K14}, an approx. 2 {\AA} continuum region around 8507.03 {\AA}, and the \ion{Ca}{2} 8542 {\AA} spectral line (see the upper part of Table \ref{tab:spectralLines}). We have also taken into account an instrumental spectral profile, or line-spread function (LSF) estimated from laser measurements by the SCIP instrumental team (Gaussian shape with a FWHM of 37.93 m{\AA}), and a theoretical PSF coming from a circular 1-meter diameter aperture together with a circular central obscuration (24.5 cm in diameter \citealt{barthol2011_sunrise}). 

The inversion process consisted of three cycles as gathered in the lower part of Tab. \ref{tab:spectralLines}. The aim of these was to get to a close initial state within a limited freedom setting (first cycle), get a detailed characterization without accounting for the PSF (second cycle), and finally include the effect of the PSF in the same inversion setting (third cycle). While it is possible to directly account for the PSF in all the cycles, experience shows that it is computationally cheaper to first get closer to the final state and include the PSF in later stages. In addition, the spatial regularization weights, both along the LOS and in the perpendicular direction, are included in the lower part of Table \ref{tab:spectralLines}. We note that the radiative transfer itself is done assuming 1.5D plane-parallel-geometry (the NLTE population calculation does not take into account the effect of radiation propagating from neighboring pixels), although the model parameters are spatially-coupled by the action of the spatial telescope PSF and by spatial regularization terms. After completing the inversions, we obtained reconstructed atmospheric properties (see Figs. \ref{fig:atm0}, \ref{fig:atm1}, and \ref{fig:atm2}) for three FOVs marked in Fig. \ref{fig:observed_fov}.

\subsection{Response Functions}
To assess the atmospheric layers sampled by the observations, we computed the response functions of the Stokes spectra to perturbations in the model parameters. They illustrate the optical-depth ranges over which the observations are sensitive to perturbations in temperature, velocity, and the magnetic-field components. These response functions are presented in Fig. \ref{fig:rf} in Appendix \ref{app:rf}. They indicate that the combined diagnostics that we included in the inversion provide sensitivity from approximately $\lg\tau_{500}\in(-5,0)$, although the exact sensitivity and vertical resolution can vary with atmospheric parameter and optical depth.

\subsection{Radiative Losses}
As a by-product of the previous process, STiC allows the resulting atmospheric models to be used to calculate the net radiative rate $Q$ for different atomic species (see for instance \citealt{morosin2022} for a description of this calculation). Under the sign convention adopted here, $Q<0$ corresponds to net radiative cooling, whereas $Q>0$ corresponds to net radiative heating. Here we have performed that calculation for \ion{Ca}{2} \citep{shine1974,uitenbroek2001} and \ion{Mg}{2} \citep{leenaarts2013a} ($Q_{\mathrm{Ca\,II}} + Q_{\mathrm{Mg\,II}}$). It is common to estimate radiative losses for \ion{H}{1} also but due to the spectral line setup in this work, a significant part of the \ion{H}{1} diagnostics opacity is outside the atmosphere in which the spectral lines we use are sensitive to and so we decided to exclude it here. This decision implies that we will estimate lower limits for these values as compared to other works, but since hydrogen contribution to radiative losses occurs mainly higher up in the atmosphere (see for instance \citealt{morosin2022}), the values discussed below are still of relevance. We discuss the derived values in Section \ref{temperature_increase} along with other properties of the model.

\subsection{Mask construction}
To further characterize the physical properties of the studied regions, we used the observed continuum intensity and the \ion{Ca}{2} wing intensity. The latter was defined as $I_{\rm wing} =\frac{1}{2}\left(I_{\rm blue}+I_{\rm red}\right)$, where $I_{\rm blue/red}=I(8542.088 \mp 0.6\ {\mathrm{\AA}})$ are the Stokes $I$ intensities at the selected blue/red-wing wavelengths. We divided each FOV into four groups: umbra (UM; continuum intensity below $2.45\times10^{-5}$ erg\,cm$^{-2}$\,s$^{-1}$\,sr$^{-1}$\,Hz$^{-1}$), umbral bright features within the pore (UBF; continuum intensity between $2.45\times10^{-5}$ and $3.28\times10^{-5}$ erg\,cm$^{-2}$\,s$^{-1}$\,sr$^{-1}$\,Hz$^{-1}$), photospheric brightening region (PB; wing intensity above $2.44\times10^{-5}$ erg\,cm$^{-2}$\,s$^{-1}$\,sr$^{-1}$\,Hz$^{-1}$, outside the combined UM and UBF mask and within five pixels of its outer boundary\footnote{At an image scale of $0.094^{\prime\prime}$ per pixel, five pixels correspond to approximately $0.47^{\prime\prime}$.}), and quiet-Sun-like region (QS; remaining pixels of the selected subfield). The ``QS'' sample is not strictly quiet Sun but comprises areas adjacent to the pore within the active region. 

The particular fixed intensity values were selected manually based on the observed morphology and were applied uniformly to all three rasters to delineate the visually identifiable continuum pore structures and \ion{Ca}{2} wing enhancement. Although the threshold selection is partly subjective, varying the thresholds by approximately 10\% does not alter the reported qualitative morphology of the masks. The spatial distribution of the masks in each FOV is depicted in Fig. \ref{fig:masks} (panels $c$, $f$, $i$). The total numbers of pixels across the three rasters are 2877, 4259, 3581, and 28763 for UM, UBF, PB, and QS, respectively. The four selected regions are mutually exclusive, and each mask was constructed independently for each raster, whereas the reported statistics combine selected pixels from all three rasters. For the third raster, the masks were constructed only over the valid observed area because the pointing jump resulted in incomplete pore coverage.

\subsection{Statistical comparison of PB and QS}
\label{sec:statistical}
To characterize the distribution of PB-QS temperature differences ($T_{PB,i}-T_{QS,j}$), we used all 3581 PB atmospheric columns and randomly selected an equal number of QS atmospheric columns. In each realization, the order of all PB columns and that of the newly selected QS columns were independently randomized before the columns were combined one-to-one. Each PB-QS combination was retained over the full optical-depth grid and used consistently to calculate the temperature difference ($\Delta T$), excess thermal energy density of the PB plasma relative to the QS reference ($\mathcal{E}_{\Delta T}$), and the differential radiative losses of the PB regions relative to the QS reference ($\Delta Q = Q_{{\rm PB},i} - Q_{{\rm QS},j}$). For each realization, we calculated the 15.85th, 50th, and 84.15th percentiles at every optical depth. Because neighboring pixels are spatially correlated through the instrumental PSF, spatial coupling, and regularization, the percentile ranges characterize the pixel-level PB-QS distributions rather than formal statistical uncertainties. The procedure was repeated 100 times using a new random QS columns selection and new random one-to-one combinations, and the resulting percentile curves were averaged across the realizations. The 50th percentile represents the median, while the 15.85th and 84.15th percentiles delimit approximately the central 68.3\% range of the sampled PB-QS difference distribution. Tests using different sample sizes and analyses performed separately for each raster yielded similar percentile curves and temperature-difference stratifications, indicating that the combined-sample result is not driven by the adopted sample size or by a single raster. We therefore present only the combined-sample result (see Section \ref{temperature_increase}).


\section{Results}
\label{results}

\subsection{Morphology and atmospheric structure of the pore}
\label{res1}

\begin{figure*}
 \centering
 \includegraphics{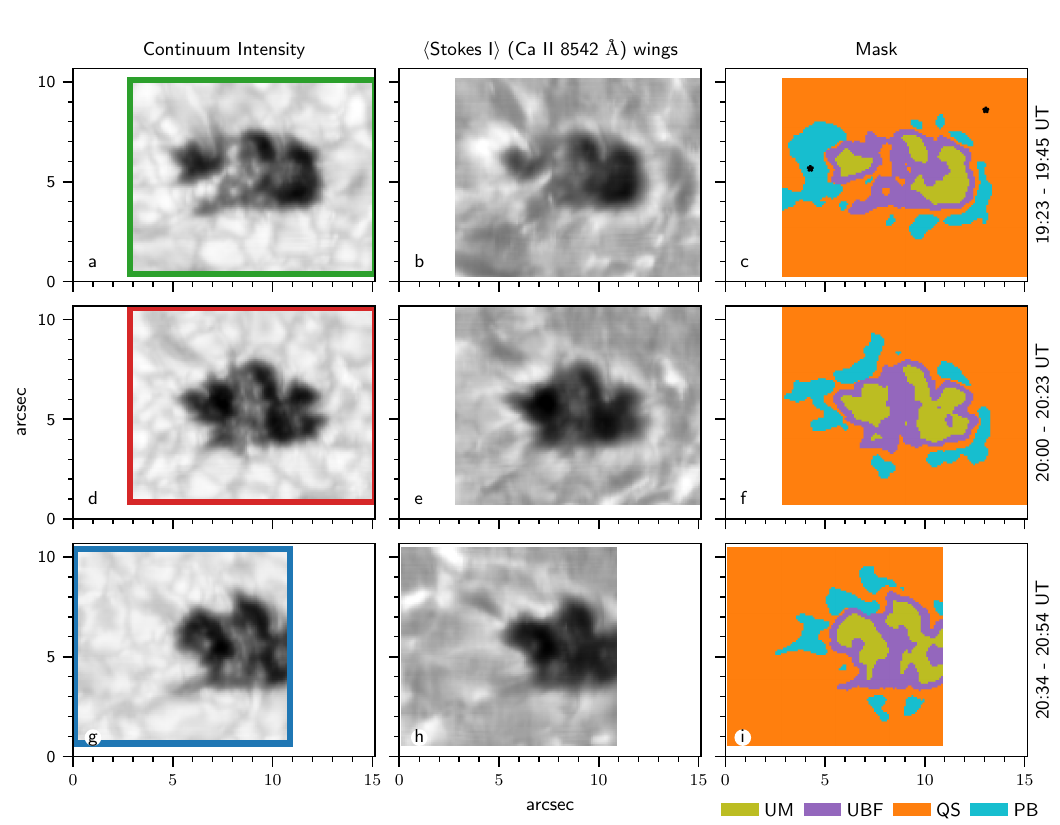}
 \caption{Continuum and \ion{Ca}{2} 8542 {\AA} wing intensity. Each row corresponds to a different raster (see right labels for the observation time). Columns display the continuum intensity, the mean Stokes $I$ in the blue and red wings of the \ion{Ca}{2} 8542 {\AA} spectral line (8542.09 $\pm$ 0.6 {\AA}), and the masks used to define the different analyzed regions (the details of the criteria applied in the main text): umbra in green (UM), umbral bright features within the pore in purple (UBF), quiet-Sun in orange (QS), and mid-photospheric brightening region in blue (PB). The colored squares in Fig. \ref{fig:observed_fov} indicate the fields of view displayed here. The black points of panel $c$ indicate the pixels used for Fig. \ref{fig:pixels}.}
 \label{fig:masks}
\end{figure*}

Figure \ref{fig:masks}, first column, shows three continuum maps of the same pore observed at three consecutive rasters with start times of 19:14, 19:51 and 20:28 UT, from first to third row, respectively. Because each raster was acquired sequentially, the resulting maps are not instantaneous and may include temporal variations along the scan direction. The structure remains coherent throughout the sequence, preserving the elongated morphology along the horizontal direction while showing substantial scan-to-scan changes in its internal structure. The internal bright structures may also be affected by evolution occurring during the raster acquisition.

In the first raster, the pore is fragmented into two parts by substantial magneto-convection in the middle part, with two filament-like features intruding from the top (as seen in Fig. \ref{fig:masks})\footnote{We want to avoid associating the bright structures we see within the pore with any type of light bridge or umbral dots because we are not entirely sure what they are}. The right part of the pore is larger and exhibits a compact and relatively homogeneous dark core, while the left segment is smaller. In the second raster, the pore displays a substantially different internal structure with multiple umbral bright features crossing its central region vertically and two main umbral cores located on either side. In the third raster, the pore appears slightly more structured, with enhanced fine-scale contrast at its boundaries. Despite these changes and the differences in the surrounding granulation among the rasters, the pore remains spatially stable and clearly distinguishable from the surrounding quiet Sun.

The second column of Fig.~\ref{fig:masks} shows the mean Stokes $I$ in the blue and red wings of the \ion{Ca}{2} 8542 {\AA} spectral line in the three different rasters, illustrating the scan-to-scan variability of the mid-photospheric intensity. Comparing it with the continuum intensity (first column), it reveals a clear transition from photospheric granulation to a more homogeneous and variable pattern at higher layers: 1- the quiet Sun loses its granular structure and exhibits brightening features, 2- a bright ring surrounds the pore and changes morphology between rasters (also detected in the valid part of the third raster), and 3- the umbral core is more spatially homogeneous, with fine umbral bright structures still visible (as in the continuum) but with reduced contrast, consistent with the behavior reported in higher atmospheric layers of light bridges and umbral dots.

\begin{figure*}
 \centering
\includegraphics{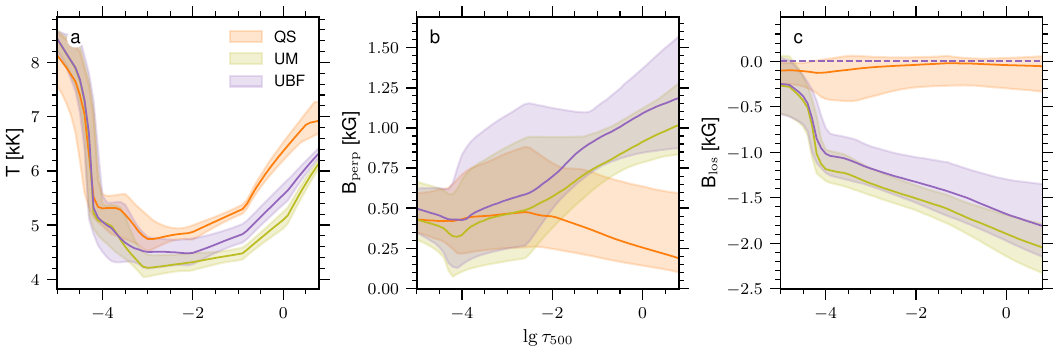}
 \caption{Comparison of the median stratifications of the inferred parameters for quiet-Sun (QS; orange line), umbral core (UM; green line), and umbral bright feature (UBF; purple line) pixels for all three rasters. Panel $a$ shows the temperature, panel $b$ the perpendicular magnetic field, and panel $c$ the LOS magnetic field. The pixels used to calculate these median stratifications are shown in the third column of Fig. \ref{fig:masks} and were selected following the criteria described in the main text. The colours of the lines are the same as used to colour the pixels in Fig. \ref{fig:masks}. The solid lines represent the median stratifications, while the shaded regions encompass the 15.85th--84.15th percentile interval, corresponding to the central 68.3\% of each distribution.}
 \label{fig:avg_temp_stratifications1}
\end{figure*}

Fig. \ref{fig:avg_temp_stratifications1} summarizes the results from the inversions. The temperature stratifications ($T$; panel $a$) show a common trend in all three regions, with the temperature decreasing toward the temperature minimum and then rising steeply toward smaller optical depths. Beyond this common behavior, the temperature stratifications reveal two distinct regimes: 1- At low to mid layers ($\lg\tau_{500}>-4$), the pore depicts the commonly found cooler UM with slightly hotter UBF and an even hotter QS. Specifically here, for these optical depths, the UM (UBF) is $\approx700$ K ($\approx400$ K) cooler than the QS. 2- Within the more limited inversion sensitivity for $\lg\tau_{500}<-4$, the retrieved temperatures of the three groups become more similar, with the pore (both UM and UBF) remaining approximately 200 K hotter.

Both magnetic components (Fig. \ref{fig:avg_temp_stratifications1} panels $b$ and $c$), along the line of sight ($B_{\rm los}$) and in the plane perpendicular to the line of sight ($B_{\rm perp}$), also exhibit clearly distinct average stratifications in these two optical depth ranges ($\lg\tau_{500}<-4$ and $\lg\tau_{500}>-4$). Both UM and UBF shows moderate $B_{\rm perp}$ values that decrease gradually with height, with the highest value in the deepest observable layer (at $\lg\tau_{500}=0$).
In contrast, the QS shows weaker transverse fields overall, especially in the photosphere, where the difference between the UM and QS values reaches $\approx0.75$ kG around $\lg\tau_{500}=0$. Toward chromospheric layers, however, the three regions show more similar values of $B_{\rm perp}$. This convergence, together with the increase of $B_{\rm perp}$ with height in the surrounding regions, is likely related to the expansion of the pore magnetic field, which forms a more inclined canopy-like configuration over the quiet-Sun-like reference.

The UM shows a strong negative polarity $B_{\rm los}$ whose magnitude increases toward deeper layers, reaching a magnitude of $\approx2.4$ kG in the photosphere. Meanwhile, the QS remains close to zero across all heights, and the UBF show intermediate values with a slightly weaker $B_{\rm los}$ than the UM. The maximum difference in magnitude in the photosphere between the values in the UM and the QS is $\approx1.6$ kG around $\lg\tau_{500}=0$.

The mean values and pixel-to-pixel standard deviations of the observed continuum intensity and inferred atmospheric parameters at three representative optical depths are provided in Table~\ref{tab:mean_parameters} in Appendix~\ref{app:models}.

\subsection{Temperature and magnetic structure of the Ca II wing brightenings}
\label{temperature_increase}

\begin{figure*}
 \centering
\includegraphics{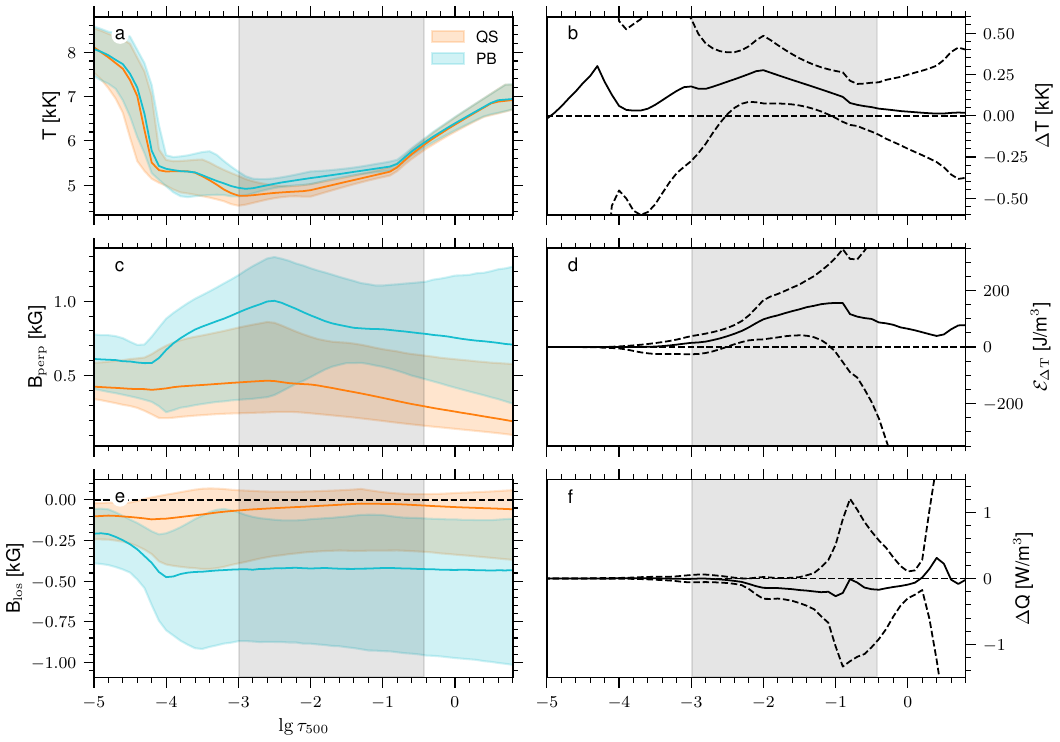}
\caption{Comparison of different physical parameters for quiet-Sun (QS; orange line), and photospheric brightening (PB; blue line) pixels for all three raster scans. Panels $a$, $c$, and $e$ show the median stratifications of the temperature, perpendicular magnetic field, and LOS magnetic field, respectively. The pixels used to calculate these median stratifications are shown in the third column of Fig. \ref{fig:masks} and were selected following the criteria described in the main text. Panel $b$ depicts the difference between the two lines represented in panel $a$ together with the standard deviation. Panel $d$ shows the excess thermal energy density of the PB plasma relative to the QS reference, and panel $f$ represents the differential radiative losses. The solid lines represent the median stratifications, while the shaded regions (dashed lines) encompass the 15.85th--84.15th percentile interval, corresponding to the central 68.3\% of each distribution in panels $a$, $c$, and $e$ ($b$, $d$, and $f$). Shaded grey areas represent the layers where QS and PB behave different.}
 \label{fig:avg_temp_stratifications2}
\end{figure*}

Figure \ref{fig:avg_temp_stratifications2} shows the distributions of $T$ (panel $a$), $B_{\rm perp}$ (panel $c$), and $B_{\rm los}$ (panel $e$) stratifications for the PB regions identified from the spectrally averaged \ion{Ca}{2} wing intensity. The figure also includes the corresponding stratification distributions for the QS reference as well as $\Delta T$ (panel $b$), $\mathcal{E}_{\Delta T}$ (panel $d$), and $\Delta Q$ (panel $f$). In all panels, the solid curves show the median, while the dashed curves show the 15.85th and 84.15th percentiles. Because $\Delta Q=Q_{\rm PB}-Q_{\rm QS}$, negative values of $\Delta Q$ indicate stronger net radiative cooling in the PB regions than in the QS reference. 

The temperature stratifications (panel $a$) of both groups (QS and PB) behave remarkably similarly at all optical depths except over the range $\lg\tau_{500}\in(-2.99,-0.43)$, highlighted in gray. This interval was defined as the optical-depth range satisfying \begin{equation} \frac{\left|P_{84.15}-P_{15.85}\right|} {\left|P_{50}\right|}<10, \end{equation} where $P_{15.85}$, $P_{50}$, and $P_{84.15}$ denote the corresponding percentiles of the PB--QS temperature-difference distribution at each optical depth. The criterion limits the percentile width of the distribution relative to the absolute median temperature difference. It also overlaps the combined sensitivity range of the photospheric diagnostics and the \ion{Ca}{2} 8542\,{\AA} line. However, the broad and overlapping response functions prevent a precise localization of the temperature enhancement within this optical-depth range. Assuming that the geometrical-height scale inferred under hydrostatic equilibrium is reasonably accurate \citep[see the discussion in][]{pastoryabar2019}, the interval corresponds approximately to heights between 100 and 400 km above the layer where $\lg\tau_{500}=0$. The response functions to $B_{\rm perp}$ also overlap this interval, supporting sensitivity to magnetic-field variations there, but not their precise localization in optical depth.

Panel $b$ shows the percentile curves of the PB-QS temperature differences obtained from the random one-to-one combinations described in Section \ref{sec:statistical}. The median PB-QS temperature difference is positive over $\lg\tau_{500}\in(-2.99,-0.43)$ and reaches approximately 300 K near $\lg\tau_{500}=-2$. The complementary mean values reported in Table~\ref{tab:mean_parameters} are consistent with this behavior: PB and QS have similar continuum intensities and temperatures at $\lg\tau_{500}=0$, whereas at $\lg\tau_{500}=-2$ the mean PB temperature is approximately 350 K higher and its mean $B_{\rm perp}$ is approximately 410 G stronger than in the QS reference. In both the PB and QS samples, $B_{\rm perp}$ increases with height, while the PB regions exhibit larger $B_{\rm perp}$ values than the QS reference over the highlighted optical-depth interval. This behavior is consistent with a field that becomes increasingly inclined as it expands away from the pore with height.

This temperature excess is inferred over an optical-depth range corresponding approximately to the middle photosphere. It is present in the combined sample from the three rasters, which were acquired over a complete observing interval of approximately 1.85 hours, and corresponds to an excess thermal-energy density relative to the surrounding plasma. To quantify this excess, we use

\begin{equation} 
  \mathcal{E}_{\Delta T} = \frac{3k_{\rm B}}{2} n_{{\rm PB},i} \left( T_{{\rm PB},i}-T_{{\rm QS},j} \right), 
\end{equation} 

\noindent where $k_{\rm B}$ is the Boltzmann constant, and $n_{{\rm lab},k}$ and $T_{{\rm lab},k}$ are the total particle number density and temperature, respectively, for a sample $k$ belonging to either the PB or QS group. The particle number density was calculated following \citet{kaithakkal2023}. By using the PB particle density for both temperature terms, this expression estimates how much the thermal-energy density increases as a result of the inferred PB--QS temperature difference. The resulting excess thermal-energy density, shown in panel $d$ of Fig.~\ref{fig:avg_temp_stratifications2}, has an average value of $\left\langle\mathcal{E}_{\Delta T}\right\rangle \approx100\ {\rm J\,m^{-3}}$  at $\lg\tau_{500}\in(-2.99,-0.43)$ and reaches approximately $170\ {\rm J\,m^{-3}}$ near $\lg\tau_{500}=-1$. Following the energetic estimate in Eq. 25 of \citet{khomenko2012}, we also calculated the equivalent magnetic-field reduction, $\Delta B$, that would provide this thermal-energy excess if the corresponding magnetic energy were converted entirely into heat. The inferred $\Delta B$ reaches approximately $20$--$30$ G, corresponding to a maximum $\Delta B/B_0$ of about $2\%$ within the selected optical-depth interval. This comparison demonstrates energetic feasibility but does not imply that such a reduction in magnetic-field strength occurred or that magnetic-energy conversion is responsible for the observed temperature enhancement.

As a final characterization of the atmospheric stratifications, we considered the difference between the combined local \ion{Ca}{2} and \ion{Mg}{2} net radiative rates in the PB and QS regions (panel $f$). The net radiative rate is given by the divergence of the radiative flux $F$:

\begin{equation}
    Q = \nabla \cdot F = \int^{\infty}_{0}\alpha_{\nu}(z) \left[S_{\nu}(z)-J_{\nu}(z)\right]\,d\nu,  
\end{equation}

\noindent where $S_{\nu}$ is the source function, $J_{\nu}$ is the angle-averaged intensity, and $\alpha_{\nu}$ is the absorption coefficient. Because $S_{\nu}$ and $J_{\nu}$ approach each other toward thermodynamic equilibrium, the calculation of $Q$ in deep atmospheric layers is particularly sensitive to the numerical convergence of the atomic populations. We therefore computed the radiative rates using a population-convergence tolerance three orders of magnitude stricter than that adopted during the inversions. The resulting $\Delta Q$ stratification indicates that the PB regions have a different radiative energy balance from the quiet-Sun-like reference, exhibiting systematically stronger net radiative cooling, with a minimum median difference of approximately $\Delta Q=-0.3$\,W\,m$^{-3}$ near $\lg\tau_{500}=-1$. If the enhanced cooling is sustained, an additional energy input would be required to maintain the inferred temperature enhancement. For completeness, the vertically integrated thermal-energy-density excess and differential radiative rate are presented in Appendix~\ref{app:integrated_energy} and Fig.~\ref{fig:enint}.

Notably, the PB temperature enhancements are not distributed continuously around the pore but appear as localized, intermediate-sized patches. Such patches are inferred in all three rasters, whose start times are separated by approximately 37 min. However, the available cadence does not allow us to determine whether individual patches persist or repeatedly re-form.

\subsection{Localized high-speed photospheric downflows} 
\label{sec:downflows} 

The inferred LOS velocity maps at $\lg\tau_{500}=0$ (panels $b$ of Figs. \ref{fig:atm0}, \ref{fig:atm1}, and \ref{fig:atm2}) show photospheric flows associated with the granular pattern surrounding the pore, together with localized downflows close to its boundary. Most of the retrieved velocities remain within the range expected for photospheric convective motions. However, in two of the three rasters (Figs. \ref{fig:atm1} and \ref{fig:atm2}), a small number of localized downflows reach $7$--$9$\,km\,s$^{-1}$. These values approach or exceed the representative adiabatic sound-speed estimate of approximately $8.2\ {\rm km\,s^{-1}}$ (calculated following Eq. \ref{eq:cs} in Appendix \ref{app:models}), indicating the possible presence of transonic or potentially supersonic photospheric flows.

Localized upflows are also visible near the pore boundary at $\lg\tau_{500}\approx-2$ (panels $f$ of Figs.~\ref{fig:atm0} and \ref{fig:atm1}), although their physical origin remains unclear.

The velocity response functions include both $\lg\tau_{500}=0$ and $-2$, although their broad optical-depth extent limits the precise vertical localization of the strongest flows. Because the observations were obtained near disk center, the LOS velocity approximately represents the local vertical velocity; however, it remains only one component of the total plasma velocity.

\section{Conclusions and discussion}

In this study, we have analyzed spectropolarimetric observations of a solar pore obtained with the SCIP instrument aboard the balloon-borne \textsc{Sunrise iii} observatory. We performed NLTE depth-stratified spectropolarimetric inversions of photospheric and chromospheric diagnostics, allowing the characterization of the thermal structure and magnetic topology of the atmosphere within and around the pore. In addition to recovering the well-established thermal and magnetic properties of solar pores, the inversions reveal spatially structured temperature enhancements surrounding the pore at mid-photospheric heights.

The inferred velocity stratifications are generally consistent with previously reported photospheric and chromospheric flow patterns \citep{sobotka2012,ermolli2017}. Localized upflows are present near the pore boundary at $\log\tau_{500}\approx-2$, although their physical origin remains unclear. In addition, two of the three rasters contain localized lower-photospheric downflows exceeding $8\ {\rm km\,s^{-1}}$ near the pore boundary. Comparable velocities have been detected around magnetic flux concentrations in plage, reaching up to $10\ {\rm km\,s^{-1}}$ at $\lg\tau_{500}=0$ \citep{buehler2015}, while compact supersonic downflows have also been reported at the outer boundaries of sunspot penumbrae \citep{vannoort2013}. Although these studies concern different magnetic environments, they demonstrate that localized supersonic downflows can occur near the boundaries of strong photospheric magnetic structures. Our inferred velocities are comparable to or exceed the representative adiabatic sound-speed estimate derived in Appendix~\ref{app:models} and may therefore indicate transonic or potentially supersonic plasma motions. Their detailed physical origin and evolution are beyond the scope of the present work.

The reconstructed depth-stratified models (see Appendix \ref{app:models}) confirm the canonical picture of pores as magnetic structures dominated by strong and mostly vertical magnetic fields in the photosphere \citep{simon1970,thomas2008}. The umbral regions are significantly cooler than the quiet-Sun-like reference at low photospheric heights, while the magnetic field strength reaches values up to $\approx2.4$ kG. The recovery of these well-established pore properties supports the reliability of the inferred atmospheric structure. 

Beyond this global photospheric characterization, the small-scale thermal and intensity structures observed in the photosphere become less pronounced toward higher atmospheric layers. A similar reduction in temperature contrast between pores and their surroundings was previously reported from \textsc{Sunrise ii} observations by \citet{solanki2017_sunrise}. That study also identified enhanced mid-photospheric temperatures near pore boundaries and a bright Ca II band around substantial portions of the pores, providing a closely related observational precedent for the structure reported here. A reduction in thermal contrast with height was also found by \citet{morosin2022} for a pore embedded in plage. In our observations, the photospheric umbral bright features are likewise no longer clearly distinguishable at the upper atmospheric layers constrained by the inversions. The present analysis extends these previous studies by recovering the magnetic and velocity stratifications over a broader atmospheric range and by estimating the thermal-energy-density excess associated with the temperature enhancement. Comparisons at a fixed optical depth do not necessarily sample the same geometrical height in regions with different opacity structures.

We have observed a structure associated with an increase in the mean Stokes $I$ in selected wavelength windows in the blue and red wings of the \ion{Ca}{2} 8542 {\AA} spectral line that has no visible counterpart in the continuum intensity maps. This enhancement in the spectrally averaged \ion{Ca}{2} 8542 {\AA} wing intensity around the pore appears as several discrete patches that collectively produce a ring-like distribution. The inferred physical parameters show that these regions are characterized by a temperature enhancement (as compared to the average temperature in the quiet-Sun-like reference) of up to $\approx300$ K and a transverse magnetic-field component approximately 500 G stronger than that of the QS reference. Although the morphology of the individual patches changes between rasters, their overall spatial distribution consistently surrounds the pore boundary defined from the continuum intensity. Their consistent spatial association with the pore across the three rasters suggests that the temperature enhancement is connected to the pore, although the temporal cadence does not allow the evolution of individual patches to be tracked directly.

The excess thermal-energy density associated with the temperature enhancement reaches approximately $170\ {\rm J\,m^{-3}}$. If this excess were supplied entirely through the conversion of magnetic energy into heat, the corresponding equivalent magnetic-field reduction would be approximately $20$--$30$ G, reaching a maximum of about $2\%$ of the inferred local field strength within the selected optical-depth interval. This comparison shows that magnetic-energy conversion is energetically feasible, but does not imply that such a reduction in magnetic-field strength occurred or that magnetic dissipation is responsible for the observed temperature enhancement.

The inferred magnetic stratifications show that $B_{\rm perp}$ increases with height in the PB and QS regions, while decreasing from the lower to the upper layers in the UM and UBF. At the same time, the maps show that the transverse field becomes more spatially extended around the pore with height. Together, these properties are consistent with the expansion of the pore magnetic field into the surrounding atmosphere, thus forming a magnetic canopy. The representative QS and PB Stokes profiles shown in Fig. \ref{fig:pixels} in Appendix \ref{app:profiles} provide complementary information about the spectral signature of the temperature enhancement. The \ion{Ca}{2} 8542 {\AA} intensity profiles are qualitatively similar to those described by \cite{delacruz2013} in the context of magnetic canopies. The PB and QS profiles have similar far-wing intensities, while the PB profile shows enhanced inner-wing intensity relative to the QS profile. Because the inner wings are sensitive to higher atmospheric layers than the outer wings, this spectral behavior is consistent with the enhancement being located above the low photosphere. The inversions place the corresponding temperature increase in layers where $B_{\rm perp}$ increases with height in the surrounding PB and QS regions and where the transverse field becomes more spatially extended around the pore. This vertical and spatial correspondence supports an association between the temperature enhancement and the expanding magnetic canopy.

This temperature enhancement does not appear to be the pore analogue of the commonly observed ring-like intensity enhancement around sunspots \citep{rast1999}. In that case, the brightness increase is readily visible in the low photosphere, whereas the temperature enhancement reported here is primarily inferred at mid-photospheric heights and has no clear counterpart in the continuum intensity maps. These differences suggest that the two phenomena may have different atmospheric origins and may therefore involve different physical processes.

Lateral radiative exchange may contribute to the observed temperature enhancement. Narrow, evacuated kG magnetic concentrations can be affected by radiation from the surrounding hot granulation \citep[e.g.,][]{nordlund2009}. Two-dimensional radiative-transfer models show that this exchange can produce heating inside magnetic flux concentrations and localized heating and cooling near their boundaries \citep{vogler2004,hasan1999}. Although the horizontal extent of the pore makes this mechanism unlikely to influence its central atmosphere efficiently, a localized contribution near its irregular outer boundary cannot be excluded. In particular, lateral radiative exchange may affect narrow or partially unresolved magnetic concentrations adjacent to the pore, where the temperature-enhanced patches are detected. Because the present 1.5D inversions neglect horizontal radiative transfer between atmospheric columns, this possible contribution cannot be quantified with the current analysis.

Alternatively, or in addition to lateral radiative heating, the expanding magnetic configuration may guide wave propagation and favor the development of strong magnetic-field gradients or current sheets. Possible heating mechanisms include wave-related dissipation, small-scale reconnection, and Ohmic or ambipolar dissipation in the partially ionized plasma \citep{ballester2018,khomenko2012,martinez2017}. However, the present observations and inversion methodology do not allow us to quantify the individual contributions of these mechanisms. For example, small-scale reconnection events and their observational signatures may remain unresolved when the spatial resolution is insufficient \citep[e.g.,][]{nobregaSiverio2024}, while the limited temporal cadence prevents us from evaluating the contribution of wave-related dissipation. The present inversion setup does not provide sufficiently reliable spatial derivatives of the vector magnetic field to estimate electric-current densities (see \citealt{meissner2026}), although alternative approaches specifically designed to recover the required derivatives have been presented by \citet{pastoryabar2021,borrero2023}. Therefore, although the inferred magnetic topology and energy budget make a canopy-associated heating process energetically plausible, the present observations do not identify the physical mechanism responsible for the temperature enhancement.

In summary, the SCIP observations reveal patchy, ring-like intensity enhancements around the pore that are associated with localized mid-photospheric temperature enhancements and a stronger transverse magnetic-field component. Their spatial correspondence with the expanding magnetic field is consistent with a canopy configuration, although the underlying heating mechanism cannot be uniquely identified. The feature is detected in all three rasters, but the available cadence does not establish whether individual patches persist or repeatedly re-form. Higher-cadence Tunable Magnetograph \citep[TuMag;][]{deltoro2025} or Sunrise Ultraviolet Spectropolarimeter and Imager \citep[SUSI;][]{feller2025} observations could help constrain their temporal evolution. The inversions also reveal localized lower-photospheric downflows reaching transonic or potentially supersonic velocities. These findings illustrate the value of solar pores as laboratories for studying the interplay between magnetic topology, convection, radiative transfer, and atmospheric energy transport.


\appendix 
\renewcommand{\thefigure}{\thesection.\arabic{figure}}
\setcounter{figure}{0}

\section{Response Functions} 
\label{app:rf}

\begin{figure*}
 \centering
\includegraphics{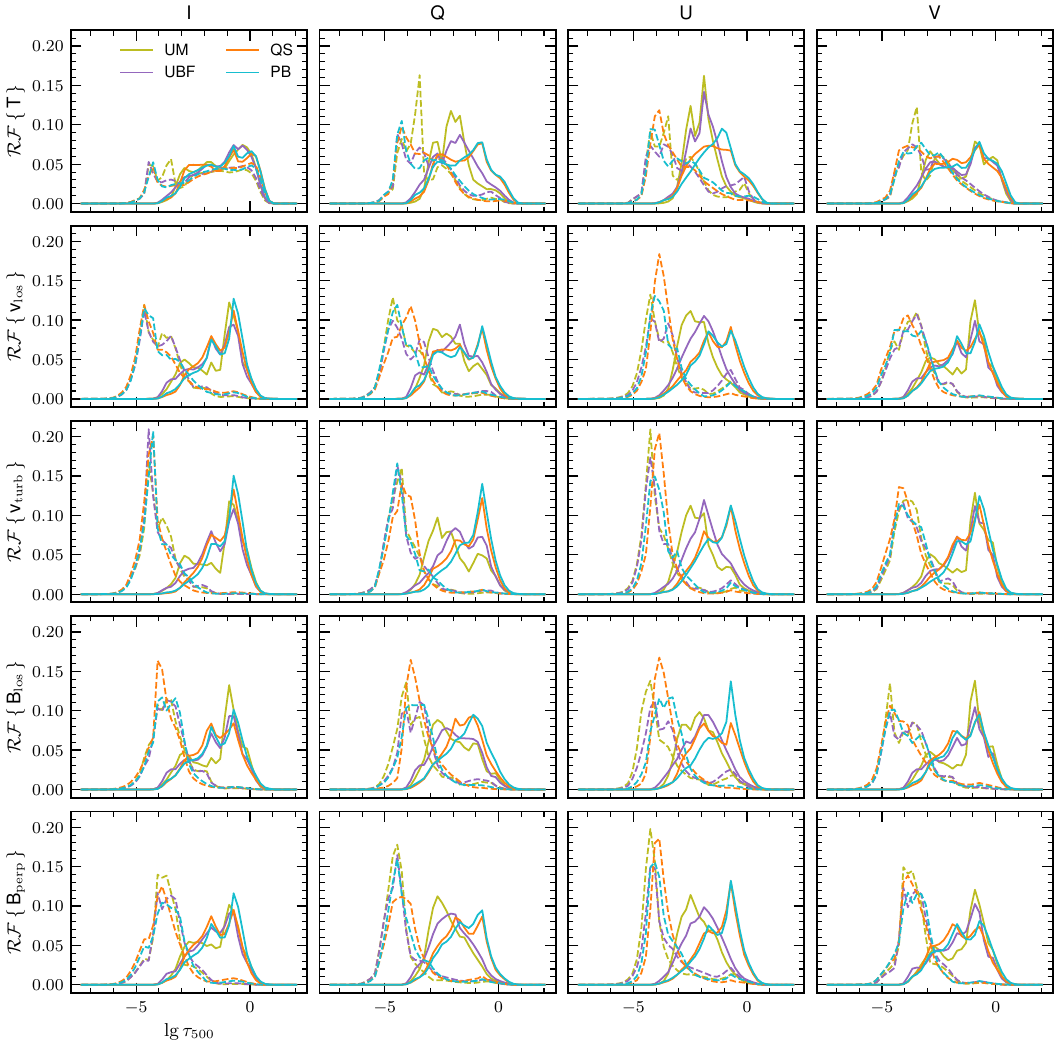}
\caption{Response functions of the inversion parameters as a function of optical depth. Rows correspond to the different atmospheric parameters (temperature, LOS velocity, microturbulence velocity, LOS magnetic field and perpendicular magnetic field), while the columns show the different Stokes parameters (I, Q, U, and V). Dashed and solid curves represent the \ion{Ca}{2} 8542\,{\AA} and \ion{Fe}{1} lines diagnostics, respectively.}
 \label{fig:rf}
\end{figure*}

Figure \ref{fig:rf} presents the response functions ($\mathcal{RF}$) of the Stokes profiles to perturbations in the atmospheric model parameters. Solid and dashed lines represent the response functions for the \ion{Fe}{1} spectral lines and continuum window and for the \ion{Ca}{2} spectral-line window, respectively. They were calculated for all pixels in the final atmospheric models. For each atmospheric parameter, spectral window, and sample (UM, UBF, PB, or QS), we determined at every optical depth the maximum absolute response among all pixels belonging to that sample. These curves therefore represent the maximum sensitivity found within each sample rather than the sample-averaged sensitivity. Overall, all groups exhibit broadly similar optical-depth coverage. The \ion{Ca}{2} spectral window is primarily sensitive over $\lg\tau_{500}\in(-5.0,-2.0)$, while the other two spectral windows probe deeper layers ($\lg\tau_{500}\in(-3.0,0.0)$).

\section{Inferred atmospheric parameters} 
\label{app:models}
\setcounter{figure}{0}

\begin{figure*}
 \centering
\includegraphics{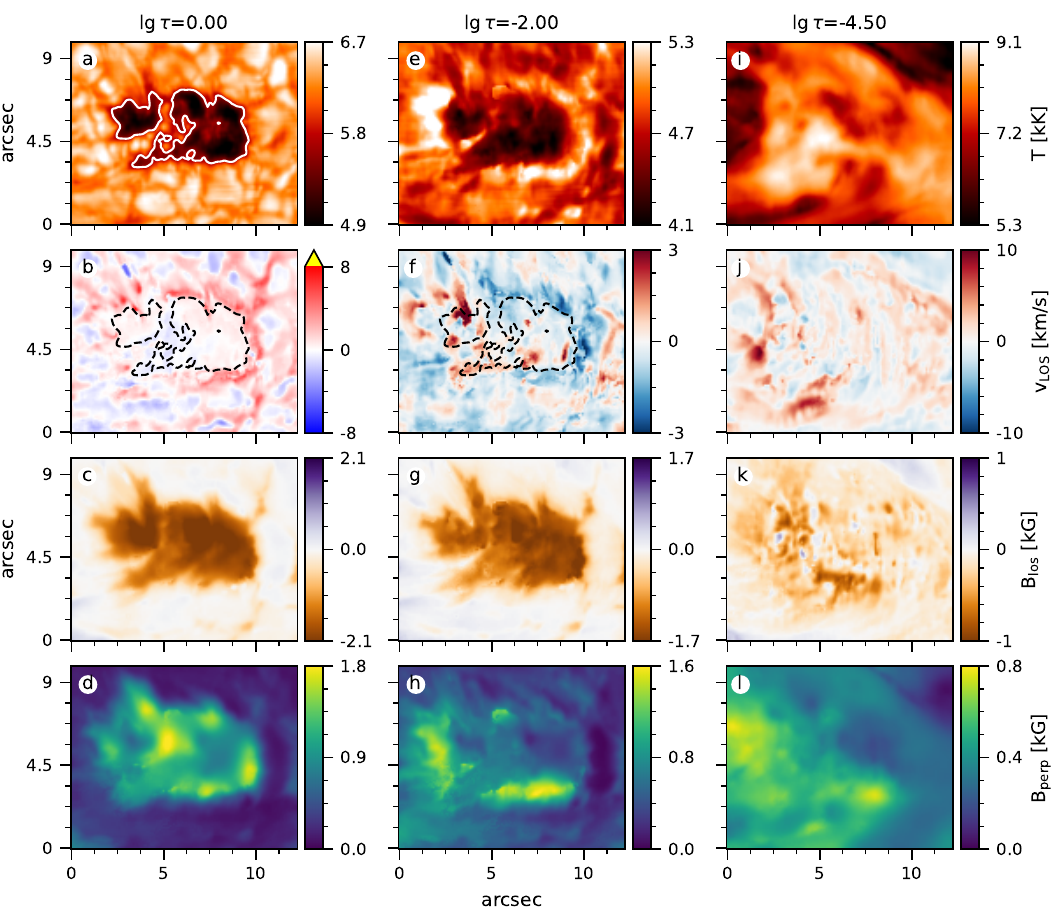}
 \caption{Inferred atmospheric parameters from the NLTE inversions for the raster observed between 19:23 and 19:45 UT. Columns correspond to different optical depths (from left to right, $\lg\tau=0$, $\lg\tau=-2$ and $\lg\tau=-4.5$) and rows to temperature in kK, LOS velocity in km/s, LOS magnetic field in kG, and perpendicular magnetic field in kG. The contour in panels $a$, $b$, and $f$ corresponds to a temperature of 5700\,K at $\lg\tau=0$.}
 \label{fig:atm0}
\end{figure*}

\begin{figure*}
 \centering
\includegraphics{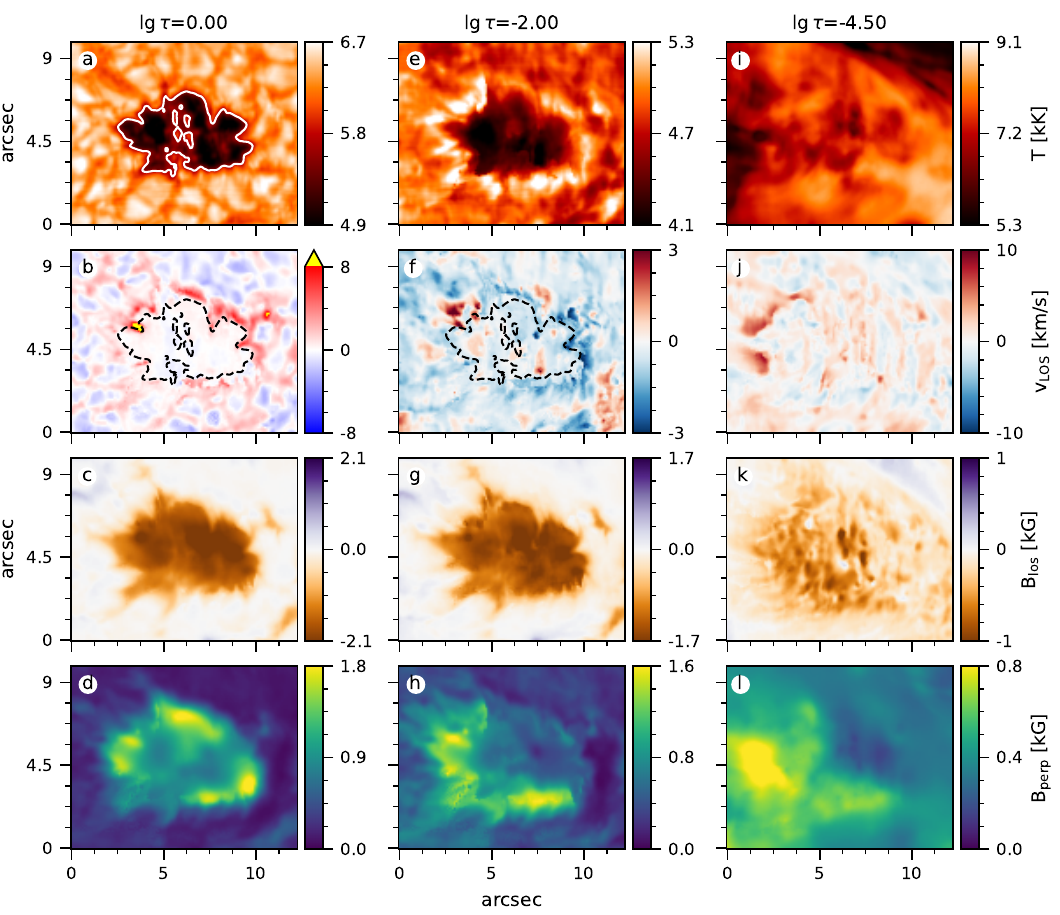}
 \caption{Inferred atmospheric parameters from the NLTE inversions for the raster observed between 20:00 and 20:23 UT. Columns correspond to different optical depths (from left to right, $\lg\tau=0$, $\lg\tau=-2$ and $\lg\tau=-4.5$) and rows to temperature in kK, LOS velocity in km/s, LOS magnetic field in kG, and perpendicular magnetic field in kG. The contour in panels $a$, $b$, and $f$ corresponds to a temperature of 5700\,K at $\lg\tau=0$.}
 \label{fig:atm1}
\end{figure*}

\begin{figure*}
 \centering
\includegraphics{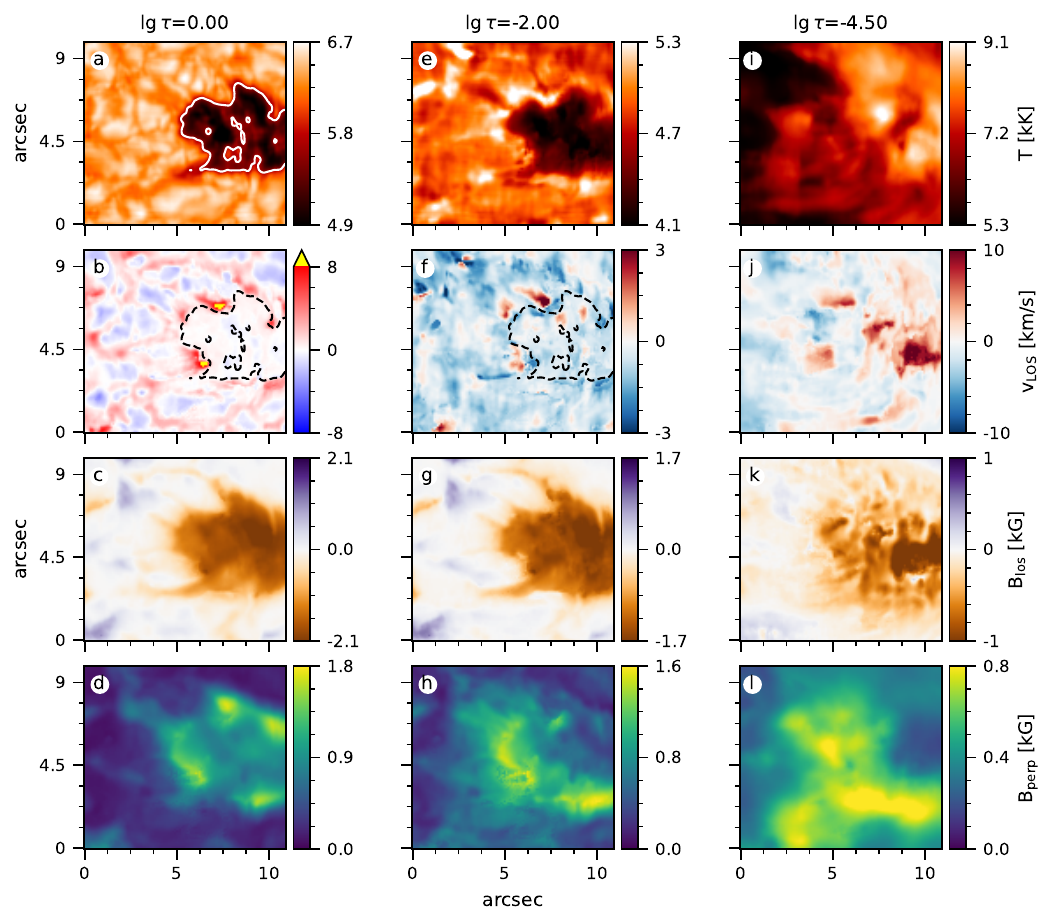}
 \caption{Inferred atmospheric parameters from the NLTE inversions for the raster observed between 20:34 and 20:54 UT. Columns correspond to different optical depths (from left to right, $\lg\tau=0.$, $\lg\tau=-2.$ and $\lg\tau=-4.5$) and rows to temperature in kK, LOS velocity in km/s, LOS magnetic field in kG, and perpendicular magnetic field in kG. The contour in panels $a$, $b$, and $f$ corresponds to a temperature of 5700\,K at $\lg\tau=0$.}
 \label{fig:atm2}
\end{figure*}

In this appendix we present the maps of the atmospheric parameters inferred from the NLTE inversions for the three SCIP rasters analyzed in this work (Figs. \ref{fig:atm0}, \ref{fig:atm1} and \ref{fig:atm2}). The figures show the spatial distribution of temperature, LOS velocity, LOS magnetic field, and perpendicular magnetic field at several optical depths spanning the photosphere and chromosphere. These figures provide a visual overview of the spatial distribution and scan-to-scan variability of the physical quantities discussed in the main text.

For completeness, Table~\ref{tab:mean_parameters} summarizes the mean observed continuum intensity and inferred atmospheric parameters for the four selected regions, combining all valid pixels from the three rasters. The reported standard deviations characterize the pixel-to-pixel dispersion within each region and should not be interpreted as uncertainties on the mean.

\begin{table}[]
  \caption{Mean observed continuum intensity and inferred atmospheric parameters for the four selected regions at three representative optical depths.}
  \label{tab:mean_parameters}
  \centering
  \begin{tabular}{lccccc}
    \colhead{Parameter} & \colhead{$\log\tau_{500}$} & \colhead{UM} & \colhead{UBF} & \colhead{QS} & \colhead{PB} \\ 
    \hline 
               & $0$    & $5.08 \pm 0.13$ & $5.56 \pm 0.18$ & $6.35 \pm 0.18$ & $6.38 \pm 0.17$ \\ 
    $T$ {[kK]} & $-2$   & $4.36 \pm 0.15$ & $4.52 \pm 0.20$ & $4.83 \pm 0.16$ & $5.18 \pm 0.16$ \\ 
               & $-4.5$ & $7.86 \pm 0.42$ & $8.04 \pm 0.57$ & $7.58 \pm 0.84$ & $7.38 \pm 0.86$ \\ 
    \hline 
                                   & $0$    & $0.33 \pm 0.24$ & $0.19 \pm 0.73$ & $0.57 \pm 1.22$  & $1.35 \pm 1.51$ \\ 
    $v_{\rm los}$ {[km\,s$^{-1}$]} & $-2$   & $0.13 \pm 0.85$ & $0.08 \pm 1.03$ & $-0.31 \pm 0.63$ & $-0.36 \pm 1.01$ \\ 
                                   & $-4.5$ & $0.50 \pm 1.28$ & $0.47 \pm 1.32$ & $0.67 \pm 1.58$  & $0.99 \pm 2.52$ \\ 
    \hline 
                         & $0$    & $-1.98 \pm 0.19$ & $-1.73 \pm 0.31$ & $-0.21 \pm 0.40$ & $-0.78 \pm 0.44$ \\ 
    $B_{\rm los}$ {[kG]} & $-2$   & $-1.48 \pm 0.14$ & $-1.27 \pm 0.27$ & $-0.18 \pm 0.34$ & $-0.43 \pm 0.41$ \\ 
                         & $-4.5$ & $-0.24 \pm 0.18$ & $-0.26 \pm 0.20$ & $-0.13 \pm 0.14$ & $-0.25 \pm 0.15$ \\ 
    \hline 
                          & $0$    & $0.95 \pm 0.16$ & $1.18 \pm 0.22$ & $0.40 \pm 0.37$ & $0.73 \pm 0.32$ \\ 
    $B_{\rm perp}$ {[kG]} & $-2$   & $0.53 \pm 0.22$ & $0.68 \pm 0.25$ & $0.48 \pm 0.27$ & $0.89 \pm 0.31$ \\ 
                          & $-4.5$ & $0.36 \pm 0.13$ & $0.41 \pm 0.13$ & $0.39 \pm 0.14$ & $0.55 \pm 0.15$ \\ 
    \hline 
    $I_{\rm cont}$  & & $2.23 \pm 0.14$ & $2.87 \pm 0.24$ & $4.11 \pm 0.27$ & $4.21 \pm 0.25$ \\
    {[$10^{-5}$ erg\,cm$^{-2}$\,s$^{-1}$\,sr$^{-1}$\,Hz$^{-1}$]} & & & & & \\
  \end{tabular} 
\end{table}

To determine whether the LOS velocities inferred near $\lg\tau_{500}=0$ are comparable to the adiabatic sound speed, we estimate the latter using:
\begin{equation}
  \label{eq:cs}
  c_s = \sqrt{\frac{\gamma k_B T}{\mu m_{\rm H}}} ,
\end{equation}

\noindent where $\gamma = 5/3$ (assuming an ideal monatomic gas), $k_B=1.380649\times10^{-23}\ {\rm J\,K^{-1}}$, $T=6400\ {\rm K}$ (approximately the average temperature obtained within the field of view for the optical depth under study), $\mu = 1.25$ (representative of the photospheric plasma), $m_{\rm H}=1.6735\times10^{-27}\ {\rm kg}$. Using the FOV-average temperature at $\lg\tau_{500}=0$, we obtain a representative adiabatic sound speed of $c_s\approx8.2\ {\rm km\,s^{-1}}$. Local values may differ because the temperature and mean molecular weight vary across the FOV. This value is used as a representative sonic reference in the velocity maps (panels $b$ of Figs. \ref{fig:atm0}, \ref{fig:atm1} and \ref{fig:atm2}). LOS velocities above this reference are regarded as potentially supersonic under the adopted assumptions.

\section{Representative Stokes profiles from a QS pixel and a PB pixel} 
\label{app:profiles}
\setcounter{figure}{0}

\begin{figure*}
 \centering
\includegraphics{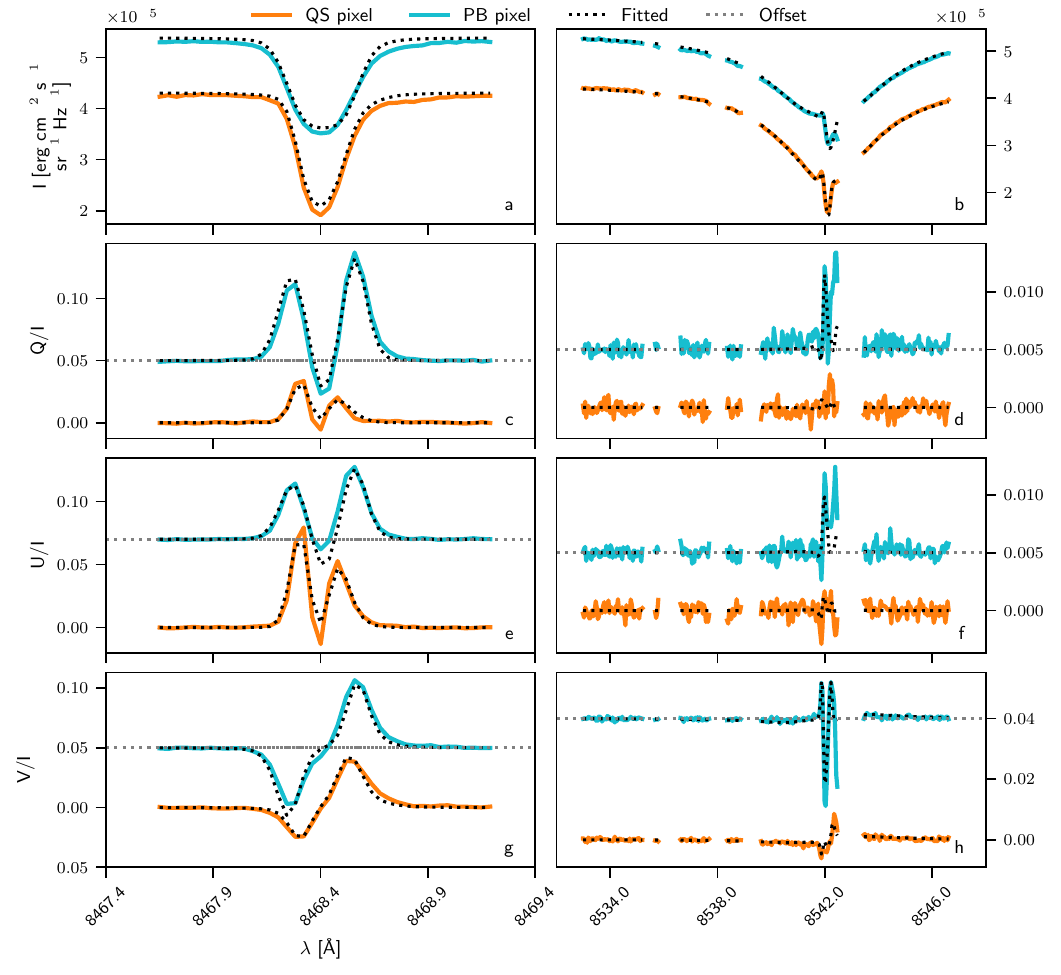}
 \caption{Representative observed and best-fitting synthetic Stokes profiles (dotted black) for QS (solid orange) and PB (solid blue) pixels. Rows show Stokes $I$, $Q$, $U$, and $V$, while columns correspond to the \ion{Fe}{1} 8468.4\,{\AA} line (left) and the \ion{Ca}{2} 8542\,{\AA} line (right). For clarity, the PB profiles are vertically offset to avoid overlap with the QS profiles. Horizontal gray dotted lines indicate their shifted zero levels.}
 \label{fig:pixels}
\end{figure*}

In this appendix we show representative Stokes profiles from pixels belonging to the quiet-Sun (QS) and photospheric-brightening (PB) regions (Fig. \ref{fig:pixels}). The displayed profiles correspond to the spectral lines included in the inversion scheme and are compared with the best-fit synthetic spectra obtained from the inferred atmospheric models. 

\section{PB integrated energy per unit area} 
\label{app:integrated_energy}
\setcounter{figure}{0}

\begin{figure}
 \centering
\includegraphics{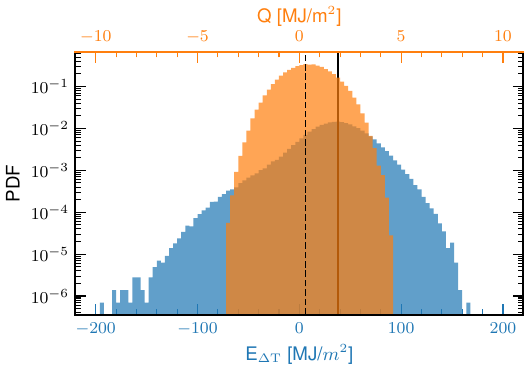}
 \caption{Distributions of the vertically integrated thermal-energy excess and differential radiative losses in the PB regions relative to the QS reference. The blue distribution shows the thermal-energy excess per unit area obtained by integrating $E_{\Delta T}$ over the geometrical-height interval corresponding to $\lg\tau_{500}\in(-2.99,-0.43)$. The orange distribution shows the radiative-energy excess over the same interval after multiplying the radiative-loss flux by the integration time per slit position. The horizontal axes for the thermal and radiative contributions are shown at the bottom and top, respectively. The solid and dashed vertical lines indicate the modes of the blue thermal-energy-excess and orange differential-radiative-energy distributions, respectively.}
 \label{fig:enint}
\end{figure}

Figure \ref{fig:enint} shows the distributions of the vertically integrated excess thermal-energy density $E_{\Delta T}$ (in blue) and differential radiative rate in the PB regions relative to the QS reference $\Delta {\rm Q}$ (in orange). Both quantities were integrated over the geometrical-height interval corresponding to $\lg\tau_{500}\in(-2.99,-0.43)$. 

To express the differential radiative rate as an energy per unit area, we multiplied the vertically integrated rate by the 10 s exposure time at each slit position. The resulting quantity therefore represents the differential radiative energy accumulated during the exposure and should not be interpreted as the total energy radiated over the unknown lifetime of the temperature-enhanced patches. The combined energy requirement has a median value of $36.4^{+26.0}_{-27.6}$\,MJ\,m$^{-2}$, where the lower and upper intervals correspond to the 15.85th and 84.15th percentiles of the PB-QS distribution.

\begin{acknowledgments}
We thank the anonymous referee for valuable comments and suggestions that improved the manuscript. The Institute for Solar Physics is supported by a grant for research infrastructures of national importance from the Swedish Research Council (registration number 2021-00169). 
ABGM acknowledges support by the Swedish National Space Agency through the Call 2023-C, project number 31004980.
APY acknowledges support from the Swedish Research Council (grant 2023-03313).
This project has been funded by the European Union through the European Research Council (ERC) under the Horizon Europe program (MAGHEAT, grant agreement 101088184).
The computations was enabled by resources provided by the National Academic Infrastructure for Supercomputing in Sweden (NAISS), partially funded by the Swedish Research Council through grant agreement no. 2022-06725, in particular at the PDC Centre for High Performance Computing (PDC-HPC) at the Royal Institute of Technology in Stockholm within project NAISS 2025/1-9.
\textsc{Sunrise iii} is supported by funding from the Max-Planck-F\"{o}rderstiftung (Max Planck Foundation), NASA under Grants \#80NSSC18K0934 and \#80NSSC24M0024 (“Heliophysics Low Cost Access to Space” program), and the ISAS/JAXA Small Mission-of-Opportunity program and JSPS KAKENHI Grant Numbers JP18H05234 and JP23K25916. This research has received financial support from the European Union’s Horizon 2020 research and innovation program under grant agreement No. 824135 (SOLARNET) and No. 101097844 (WINSUN) from the European Research Council (ERC). It has also been funded by the Deutsches Zentrum für Luft- und Raumfahrt e.V. (DLR, grant no. 50 OO 1608). The Spanish contributions have been funded by the Spanish MCIN/AEI under projects RTI2018-096886-B-C5, and PID2021-125325OB-C5, and from “Center of Excellence Severo Ochoa” awards to IAA-CSIC (SEV-2017-0709, CEX2021-001131-S), all co-funded by European REDEF funds, "A way of making Europe”
This work has made use of the VALD database, operated at Uppsala University, the Institute of Astronomy RAS in Moscow, and the University of Vienna.
This research has made use of the Astrophysics Data System, funded by NASA under Cooperative Agreement 80NSSC25M7105.
Microsoft 365 Copilot was used under human supervision to assist with English-language editing, grammatical corrections, paragraph organization, and improvements to textual clarity. All suggestions were reviewed and revised by the first author as necessary. The tool was not used to generate or analyze data, create figures, or determine the scientific results or conclusions. Responsibility for the final manuscript lies entirely with the first author.
\end{acknowledgments}


\bibliography{sample701}{}
\bibliographystyle{aasjournalv7}




\end{document}